\documentclass{article}

\usepackage{makeidx}         
\usepackage{graphicx}        
\usepackage{multicol}        
\usepackage[bottom]{footmisc}

\usepackage{subfigure}
\usepackage{amsmath}
\usepackage{algpseudocode}
\usepackage{algorithm}

\makeindex   

\begin{document}

\title{Epidemics on a stochastic model of temporal network}
\author{Luis E. C. Rocha \and Adeline Decuyper \and Vincent D. Blondel \\
Department of Mathematical Engineering \\
Universit\'e catholique de Louvain, Louvain-la-Neuve, Belgium\\
Luis.Rocha@uclouvain.be}
\date{20 March 2012}

\maketitle

\section{Introduction}
\label{intro}

Individual contacts between people serve as pathways where infections may propagate. Characterizing and modeling these contacts is, therefore, fundamental to unveil potential infection routes, to understand the emergence of epidemics, and to control or avoid the epidemics~\cite{BBV08}. One way to represent the patterns of contacts is through networks~\cite{BBV08, N10}. The network structure captures the heterogeneity of individual behavior and goes beyond random mixing where everyone can interact with everyone else. While static structures have been the common modeling paradigm, recent results suggest that temporal structures play different roles to regulate the spread of infections~\cite{RLH11, SKL10, SVB11} (or infection-like dynamics, as in information spreading~\cite{KKP11}). This has shifted the research attention to the understanding of such changing structures~\cite{HS12}. On temporal networks a particular vertex is active only at certain moments and inactive otherwise such that a contact is not continuously available~\cite{RLH11, KKP11, HS12}. The temporal order of contacts restricts the infection routes between vertices in comparison to non-temporal networks~\cite{TSM10,PS11}. The time between two consecutive vertex-activation events is not necessarily uniform, but in many datasets it follows heterogeneous patterns (e.g.\ bursts), in particular it does so on several networks of relevance to spread of infections, as for example, sexual~\cite{RLH10}, proximity~\cite{SKL10, SVB11}, or communication contact networks (electronic virus)~\cite{IE09, KKP11}.

Previous research is mostly based on data-driven studies in which the evolution of the epidemics is contrasted between the original temporal sequence and randomized versions, where the original time-stamps in the edges are shuffled to destroy the temporal correlations~\cite{RLH11,KKP11}. While this approach is adequate for many purposes, it is difficult to study the sole effect of a particular temporal constraint. Other topological structures (e.g.\ degree distribution, degree-degree correlation or community structure) can be present and also contribute to regulate the infection spread. Some studies suggest that heterogeneous inter-event time implies on a slow decay time of the prevalence during an infection dynamics~\cite{VRL07} and slowdown of the spreading dynamics in the context of communication networks~\cite{IE09, KKP11}. The opposite effect, i.e.\ the speed up of the infection growth due to broad inter-event activation times, is observed in a sample of sexual contacts network~\cite{RLH11}. To improve the understanding of such contrasting behavior, in this chapter, we present a simple and intuitive stochastic model of a temporal network and investigate how epidemics co-evolves with the temporal structures, focusing on the growth dynamics of the epidemics. The model assumes no underlying topological structure and is only constrained by the time between two consecutive events of vertex activation, hereafter called vertex inter-event time.

The network model consists on random activations of a vertex according to a pre-defined vertex inter-event time distribution. The vertices active at a given time are randomly connected in pairs during one time unit. The link is then destroyed and the vertices set to the inactive state. The infection event occurs through this link if one of the vertices is in an infective state. We study this model by using a susceptible infective dynamics with one (SI) and two (SII) infective stages. The first dynamics is motivated for being an upper limit case, where once infected, the vertex continues infecting at every contact~\cite{H00}. The second dynamics is more realistic and corresponds to a model of HIV spreading including an acute (high infectivity) and chronic (low infectivity) stages of infection with different periods~\cite{HAF08}. If the second stage is set to zero in the SII model, we recover the susceptible infected recovered (SIR) dynamics~\cite{H00}.

To study our model, we compare the effects of two vertex inter-event time distributions on the spread of an infection. We consider the geometric\footnote{The geometric is a discrete time equivalent to the exponential distribution.} (corresponding to uniform probability of being active) and the power-law (to reproduce heterogeneous inter-event patterns) distributions. The main observation is that the speed of the infection spread is different for both cases but the differences depend on the stage of the epidemics. In comparison to the homogeneous scenario, the power law case results in a faster growth in the beginning but turns out to be slower after a certain time, taking several time steps to reach the whole network. 

The chapter is organized in $4$ sections. We introduce the stochastic model and discuss some structural properties of the resulting network in Section~\ref{netmodel}, and introduce and describe the epidemics models in details in Section~\ref{sprproc}. The results of the SI epidemics on the networks following the different vertex inter-event time distributions are presented and discussed in Section~\ref{resul} together with the results of the SII model. We conclude the chapter in Section~\ref{concl} highlighting the major contributions.

\section{Network Model}
\label{netmodel}

In this section, we describe the stochastic model of the temporal network and discuss some key structural properties of the evolving network.

\subsection{Network evolution}

The network is dynamic in the sense that vertices are active only at certain moments in time and inactive otherwise. As a consequence, edges also appear and disappear throughout the network evolution. In our model, whenever a vertex becomes active, it is randomly connected to another active vertex. The corresponding edge remains available during one time step, and is destroyed afterwards (see Algorithm~\ref{algnet}).

\begin{algorithm}
\caption{Network model}
\label{algnet}
\begin{algorithmic}
\State Initialization step:
\For{$i = 1 \to N$} 
\State $T_{next_i} \gets \text{number drawn from } D_{\text{pow}}$
\EndFor
\State Network Evolution: 
\For{$t=1,2,...$}
\State $T_{next} \gets T_{next}-1 $
\State $ V \gets $ all nodes $v$ such that $T_{next_v} =0$
\State \If {$\left| V \right|$ is odd }
\State select one node $u$ at random from $V$
\State $T_{next_u} \gets 1$
\State $V \gets V\backslash u $ 
\EndIf
\State \While{$V \neq \emptyset$} 
\State select two nodes at random $u,v \in V$ 
\State make a link between $u$ and $v$
\State $T_{next_u} \gets$ number drawn from $D_{\text{pow}}$
\State $T_{next_v} \gets$ number drawn from $D_{\text{pow}}$
\State $V \gets V\backslash \left\lbrace u,v \right\rbrace$
\EndWhile
\EndFor
\end{algorithmic}
\end{algorithm}

The activation time is a fundamental property of our model and depends on the system of interest. It has been observed in different contexts that vertices are not active uniformly in time but often follows different waiting times, i.e.\ the inter-event time $\Delta T_{\text{V}}$ between two vertex activations is not uniform but follows heterogeneous patterns. In particular, this inter-event time on sexual and proximity contacts, email and cell phone communication patterns are well described by power-law like $\Delta T_{\text{V}}$ distributions~\cite{RLH10,SKL10, SVB11,IE09, KKP11}. A power-law $\Delta T_{\text{V}}$ distribution means that there are bursts, i.e.\ trains of activations followed by periods of inactivity of various lengths. There are different theories trying to explain such behavior in the context of communication activity. One theory is the priority queuing model~\cite{B05} and the other is based on a nonhomogeneous Poissonian process including periodic activity patterns~\cite{MSM08}.

\begin{figure}
\centering
\includegraphics[scale=0.6]{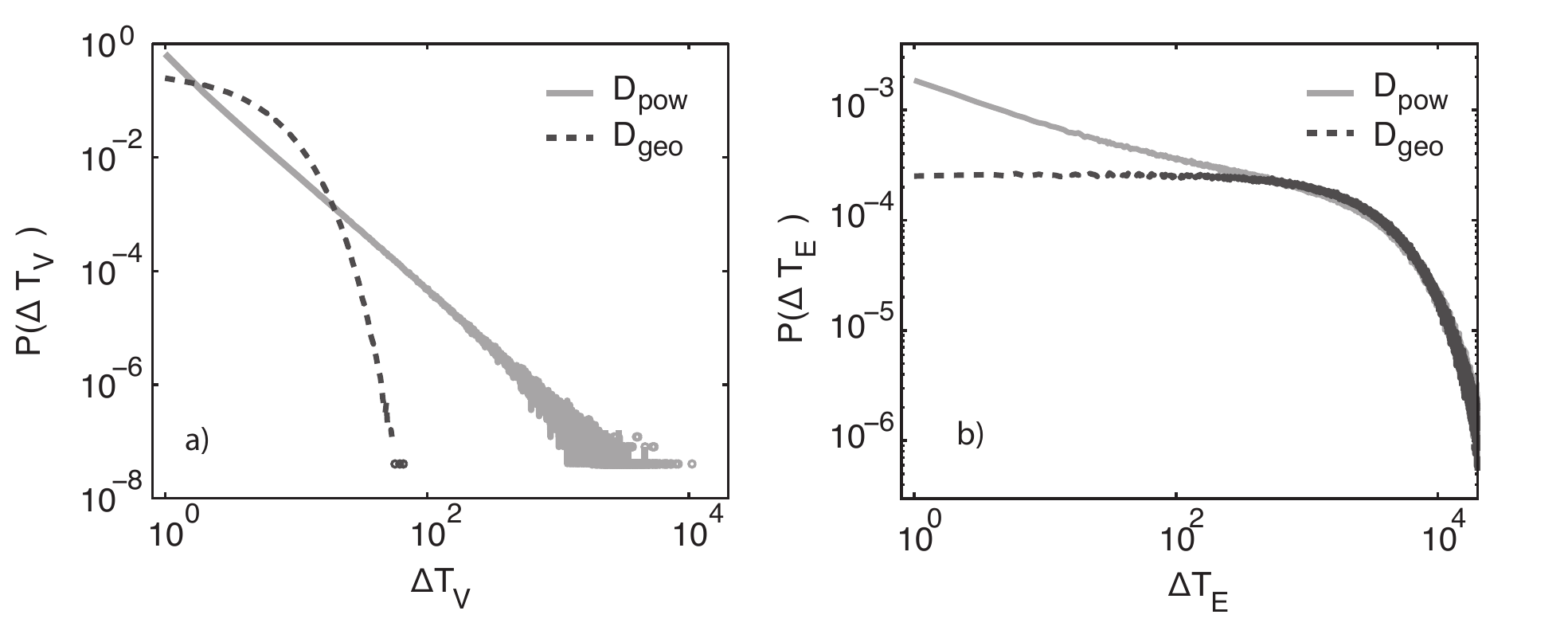}
\caption{Probability of $\Delta T$, the inter-event time between subsequent activations, for the power-law with cutoff (full line) and the geometric cases (dashed line) on the a) vertices and b) edges. The axes are in log-scale.}
\label{fig1}       
\end{figure}

To simulate this characteristic in our model, for each active vertex at time $t'$, we sample the next activation time $t_{\text{next}}$ from a power-law (with cutoff) inter-event time distribution (eqn.~\ref{eqn1}, hereafter referred to as $D_{\text{pow}}$) such that $\Delta T_{\text{V}} = t_{\text{next}} - t'$. Note that this is equivalent of saying that the probability of being active at time $t_{\text{next}}$, given that the vertex was active at time $t'$, decreases as $t_{\text{next}}$ gets larger and this decrease follows a power-law. Since we have a stochastic model, the cutoff is necessary to avoid that too large $t_{\text{next}}$'s are selected during the evolution of the network. Since the epidemics dynamics occurs within few time steps, we set the cutoff to a reasonable value ($\lambda_1 = 5\cdot10^{-4}$) so that the exponential cutoff does not affect the dynamics and the power-law regime prevails. In line with empirical observations of the inter-event time (e.g.\ refs.~\cite{RLH10,KKP11}), we set alpha to $2$ (Fig.~\ref{fig1}a). Random initial activation times $t_{\text{next}}$ are also selected from the distribution to all vertices, and to remove transient oscillations we let the network evolves during $1,000$ time steps before the analysis (see Algorithm~\ref{algnet}).

\begin{equation}
\label{eqn1}
P(\Delta T) \propto \Delta T^{-\alpha} e^{-\lambda_1 \Delta T},
\end{equation}

We compare the effects of the heterogeneous inter-event time with a null model where the probability $p$ of a vertex being active is constant (Poissonian process). This constant probability gives an exponential distribution of inter-event times (eqn.~\ref{eqn2}, hereafter referred to as $D_{\text{geo}}$) where $p=1/\lambda_2$ ($\lambda_2$ is the mean inter-event time and set to be the same value as obtained from the $D_{\text{pow}}$ case, i.e.\ $\lambda_2 = 4.024$ for $\alpha = 2$). To be more accurate in the simulations, we use a geometric distribution since we are working in discrete times.

\begin{equation}
\label{eqn2}
P(\Delta T) \propto e^{-\lambda_2 \Delta T},
\end{equation}

To generate random samples following the geometric distribution, one can use the inverse transform sampling, rounding down the value for the nearest integer~\cite{CSN09}. This function is typically available in math libraries for most programming languages. The case of the $D_{\text{pow}}$ distribution is a bit trickier and we use a method similar to the one proposed on ref.~\cite{CSN09}, shortly described by Algorithm~\ref{plcogen}. 

\begin{algorithm}
\caption{Random values using a power-law with exponential cutoff}
\label{plcogen}
\begin{algorithmic}
\While{$x$ is rejected}
\State $r \gets $ random number drawn uniformly from the interval $\left[ 0,1 \right)$
\State $x \gets \lfloor \left( x_{\text{min}}-\frac{1}{2} \right) \left(1-r \right)^{-1/\left( \alpha -1 \right)}+\frac{1}{2} \rfloor$
\State $\rho \gets e^{-\lambda x}$, probability of acceptance of $x$
\State $p \gets  $ random number drawn uniformly from the interval $\left[ 0,1 \right)$
\If {$p< \rho$}
\State \Return $x$
\Else
\State $x$ is rejected
\EndIf
\EndWhile
\end{algorithmic}
\end{algorithm}

We show on Figure~\ref{fig1}a numerically generated random samples using the methods described above for the two distributions $D_{\text{pow}}$ and $D_{\text{geo}}$. It is visible that the exponential cutoff in the power-law has little or no effect for $\Delta T_{\text{V}}$ values smaller than $300$, but still the maximum values can reach at about $\Delta T_{\text{V}} = 10,000$, while the maximum values of the geometric distribution are not larger than $70$.

\subsection{Characteristics of the temporal network}

\begin{figure}
\centering
\includegraphics[scale=0.5]{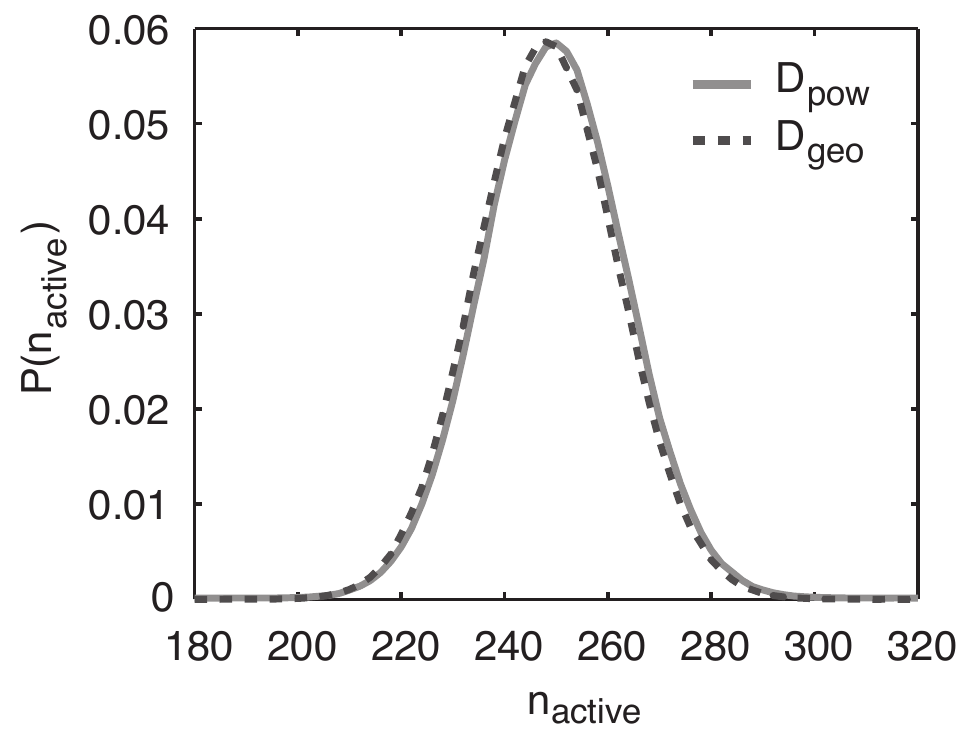}
\caption{Probability of having $n_{\text{active}}$ active vertices per time step in a network with a total of $1,000$ vertices.}
\label{fig2}       
\end{figure}

The proposed model creates evolving networks without assuming any information about the network topology. After few time steps, the majority of the vertices have been in contact to every other vertex at least once, which means that the vertex-degree is approximatelly $N_{\text{V}}$. It is, therefore, more meaningful to highlight that the distribution of the total number of contacts has a characteristic value, although the spread is larger for $D_{\text{pow}}$ in comparison with the case of $D_{\text{geo}}$. In other words, the vertices have a characteristic number of contacts with the same partner. The number of active vertices per time step $n_{\text{active}}$ is similar for both cases and for simulations using a total of $1,000$ vertices, $n_{\text{active}}$ oscillates about a mean value of about $250$ vertices (Fig.~\ref{fig2}).

The dynamics of the specific vertex inter-event times has different consequences on the edge inter-event time $\Delta T_{\text{E}}$, i.e.\ the time between the subsequent activation of the same edge. The power-law inter-event time on the vertices results in a distribution of $\Delta T_{\text{E}}$ following a power-law functional form during a long interval, a characteristic also observed on empirical analysis of sexual contacts~\cite{RLH10} and cell phone communication~\cite{KKP11}. As expected, the uniform activation probability of the $D_{\text{geo}}$ case gives a constant probability for the $\Delta T_{\text{E}}$ distribution (Fig.~\ref{fig1}b).

\section{Epidemics models}
\label{sprproc}

In this section, we describe epidemics models and present our simulation results for the spread of infections in the evolving network structure.

\subsection{SI and SII models}

To investigate the impact of the heterogeneous vertex inter-event time on the spread of infections, we simulate two complementary epidemics models on the evolving network. The first model, susceptible infective (SI), mimics a scenario where the vertices remain infective after an initial infection. Although this model is unrealistic to model actual infections, it is a good case study because it corresponds to an upper bound of infection or a worst-case scenario. To study the propagation of a hypothetical infection with a more realistic model, we run simulations for the susceptible infective model with two infective stages (SII). This model is adequate to represent the spread of HIV where an individual is highly infective during the initial months after a primary infection (acute stage), followed by a chronic stage of low infectivity.

In both epidemics models, all vertices start the dynamics at a susceptible state, except for one random vertex in the infected state. A vertex becomes infective after a contact with another infected vertex with probability $\beta$ (per-contact infection probability or transmissibility). For the SI model, the states do not change after an event of infection, but for the SII, the probability of vertex A infecting vertex B is $\beta_1$ in the initial $T_1$ time steps after the first infection of $A$, and $\beta_2$ after $T_1$ time steps. Assuming that one time step corresponds to one day, which is reasonable approximation for sexual contact networks, we set $T_1=90$ days, which is the average time of the acute stage for HIV, and following typical values, we use $\beta_1/\beta_2=10$~\cite{KJW97, HAF08}. We run the simulations on 100 network ensembles with 10 random infection seeds for each and thus we average our results over $1,000$ simulations.

\subsection{Mean-field SI epidemics for the geometric case}

In this section, we discuss the case of the temporal network where the vertex activation follows a Poissonian process that can be described by a simple mean-field approximation. A discrete time SI epidemics evolving within a population of vertices homogeneously mixed can be described by the set of equations~(\ref{SIevol})~\cite{A94}.

\begin{equation}
\label{SIevol}
\left\lbrace \begin{aligned}
S_{t+1} &=& S_t \left( 1- \frac{\gamma \Delta t}{N} I_t \right) \\
I_{t+1} &=& I_t \left( 1+ \frac{\gamma \Delta t}{N} S_t \right)
\end{aligned}
\right. 
\end{equation}
where $S_{t}$ and $I_{t}$ represent respectively the number of susceptible and infected vertices in the network at time $t$, $N$ is the total number of vertices in the network ($N=S_t+I_t$ is constant for all times), and $\gamma$ is the average number of infectious contacts an infected individual makes per unit time interval. The initial condition is given by $I_0 >0$ and $S_0=N-I_0$. 

The system of equation~(\ref{SIevol}) describes the SI dynamics on our geometric case if we set $\Delta T = 1$ and $\gamma = \lambda \beta$, where $\lambda$ is the average number of contacts per vertex per time step, and $\beta$ is the probability of per-contact infection between a susceptible and an infected vertex. Since each active vertex makes exactly one contact during a given time step, the average number of contacts per vertex per time step is, consequently, the probability for a given vertex to be active at a given time step (eqn.~\ref{lambda}).

\begin{equation}
\label{lambda}
\lambda = Prob(\text{a vertex is active at time } t) = \frac{\langle n_{\text{active}}\rangle}{N}
\end{equation}

\section{Results}
\label{resul}

In this section we present the results on the co-evolution of the epidemics model and the network structure, and discuss the impact of the heterogeneous vertex inter-event time on the spread of infections.

\subsection{Effect of the inter-event time on the SI dynamics} 

\begin{figure}[ht]
\centering
\includegraphics[scale=0.65]{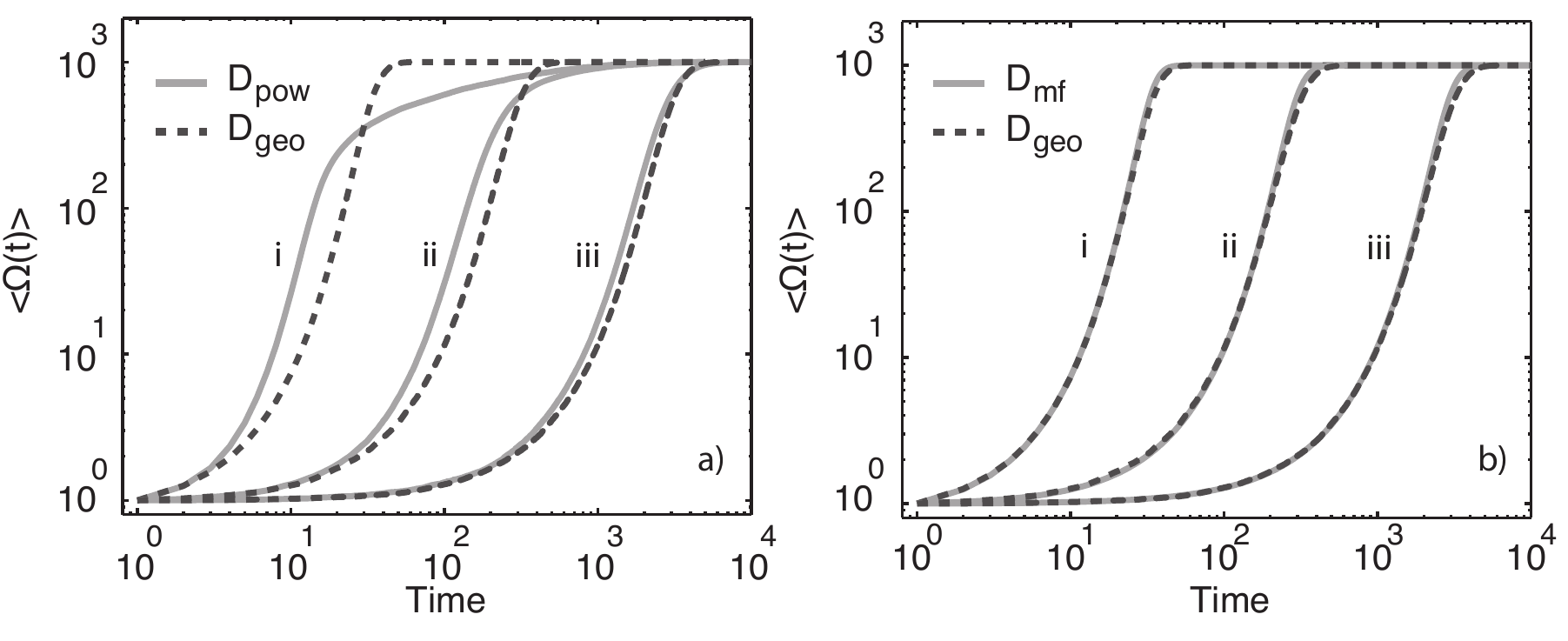}
\caption{The average number of infected vertices, $\langle \Omega (t) \rangle$, for 3 probabilities of infection, $\beta=$ 1 (i), 0.1 (ii), and 0.01 (iii), for a) power-law $D_{\text{pow}}$ and geometric $D_{\text{geo}}$ cases, and b) mean-field $D_{\text{mf}}$ and geometric $D_{\text{geo}}$ cases. The axes are in log-scale.}
\label{fig3}
\end{figure}

To obtain a global picture of the propagation of the infection, we show on Figure~\ref{fig3}a the average number of infected vertices per time, $\langle \Omega(t) \rangle$. The figure contrasts the results for the different distributions of inter-event time $D_{\text{pow}}$ and $D_{\text{geo}}$. For all studied parameters, the infection grows faster in the initial period for the case of $D_{\text{pow}}$ in comparison to the case of $D_{\text{geo}}$ but slows down afterwards, such that it takes longer to infect all vertices in the case of a power-law distribution. The difference in the growth patterns is weaker for smaller values of $\beta$. During the initial interval before the inflexion point, for both inter-event cases, the infection grows exponentially but at different rates.

Figure~\ref{fig3}b shows the evolution of $\langle \Omega(t) \rangle$ for the mean-field ($D_{\text{mf}}$) approximation and for the numerical simulation of the stochastic model using the geometric vertex inter-event time. For the mean-field case, we set an initial condition $I_0=1$ and $n_{\text{active}} = 249.75$ (a value obtained from the numerical simulation, see Fig.~\ref{fig2}) which gives $\lambda_2 = \frac{249.75}{1,000}\sim0.25$ in the case of a network with $1,000$ vertices. The theoretical and numerical curves closely overlap for most of the interval of study, with a little mismatch when almost all the network is infected.

\begin{figure}[ht]
\centering
\includegraphics[scale=0.65]{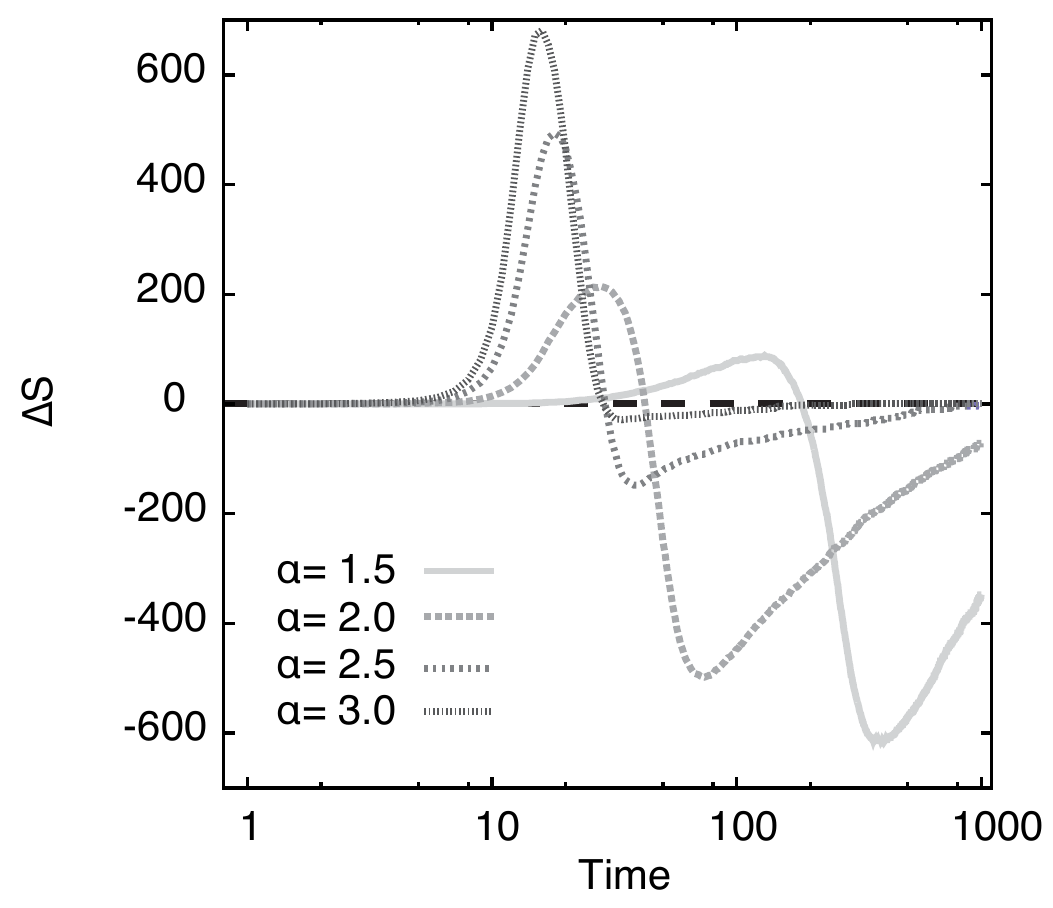}
\caption{Difference between the number of infected individuals if the vertex inter-event time follows a power-law distribution and the number of infected individuals if the vertex inter-event time follows a geometric distribution, $\Delta S = \langle \Omega(t)_{\text{pow}} \rangle -\langle \Omega(t)_{\text{geo}} \rangle$. The letter $\alpha$ corresponds to the exponent of the power-law, and the geometric version always uses the same average inter-event time as the power-law. The x-axis is in log-scale.}
\label{fig4}
\end{figure}

The speed up in the early stage and the slowdown in the final stage of the infection growth for the power-law case in comparison to the geometric case is observed for other values of $\alpha$, i.e.\ the exponent of the vertex inter-event time distribution (eqn.~\ref{eqn1}). Figure~\ref{fig4} shows the difference in the number of infected vertices in the power-law case, $\langle \Omega(t)_{\text{pow}} \rangle$, and the number of infected vertices in the geometric case $\langle \Omega(t)_{\text{geo}} \rangle$ during the simulation of $1,000$ time steps. Positive values indicate that the power-law case infects more vertices than the geometric one, and negative values indicate the opposite behavior. Independently of the exponent, we observe that during a significant interval, the power-law case leads to a faster infection growth, and this interval increases with decreasing $\alpha$.

\begin{figure}[ht]
\centering
\includegraphics[scale=0.6]{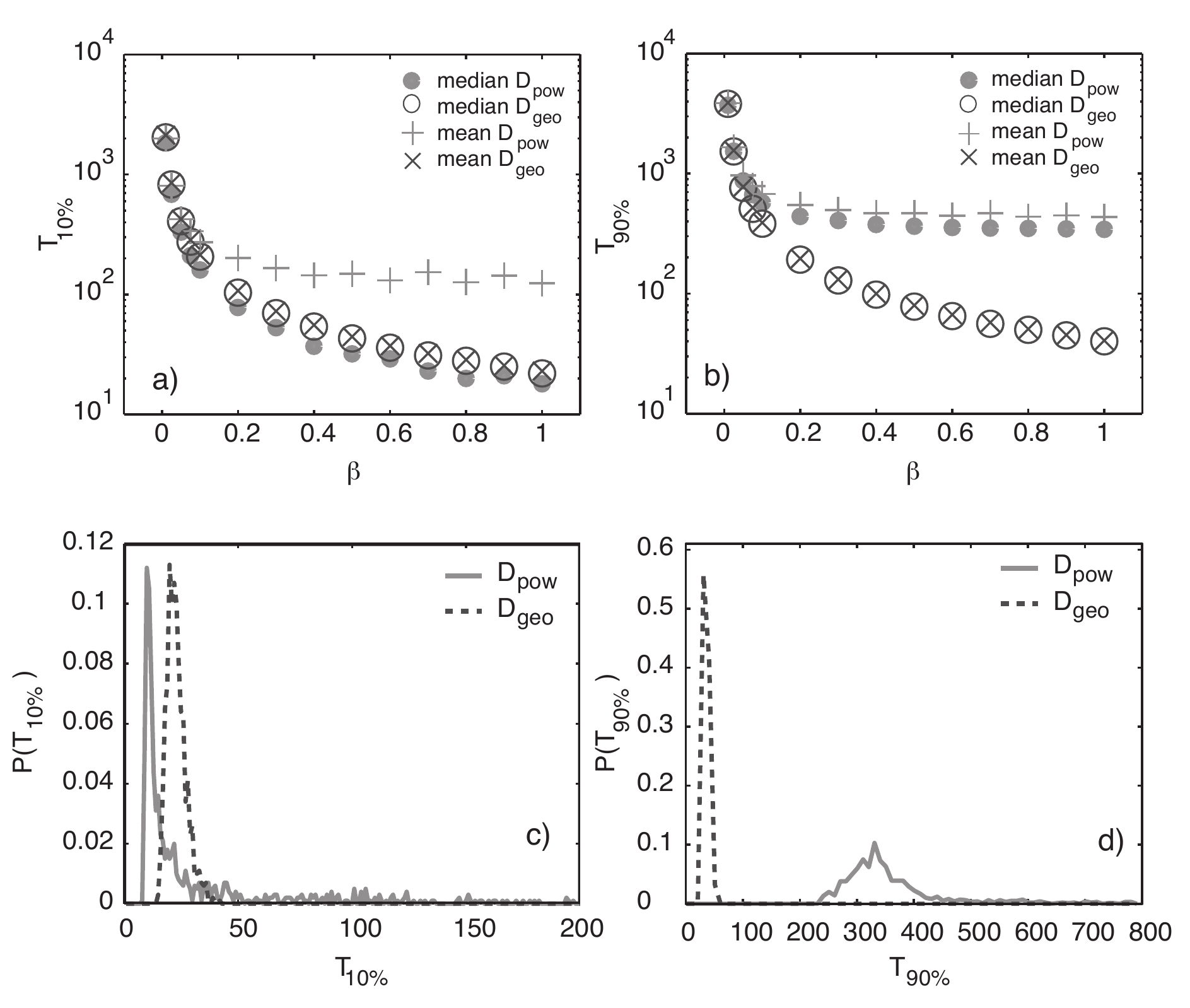}
\caption{The per-contact infection probability $\beta$ versus the mean and median time for the infection to reach a) $10\%$ and b) $90\%$ of the network for both $D_{\text{pow}}$ and $D_{\text{geo}}$ cases. In case of $\beta=1$, the probability of c) $10\%$ and d) $90\%$ of the vertices being infected after, respectively, $T_{10\%}$ and $T_{90\%}$ time steps. The y-axes in a) and b) are in log-scale.}
\label{fig5} 
\end{figure}

To better understand and characterize the differences in the growth in the early and final stages, we plot the number of time steps necessary to reach $10\%$ ($T_{10\%}$, Fig.~\ref{fig5}a) and $90\%$ ($T_{90\%}$, Fig.~\ref{fig5}b) of the vertices. We compare how the mean and the median values of this measure vary with $\beta$ for each vertex inter-event case. The median is a better statistics for non-symmetric broad distributions, whereas both the median and the mean are adequate statistics for symmetric distributions. Since the distribution of $T_{10\%}$ has no characteristic value for the power-law case (see Fig.~\ref{fig5}c and discussion bellow), we use both statistics to characterize the distributions and to compare the diverging effects. Figure~\ref{fig5}a shows that the mean and median for both $D_{\text{pow}}$ and $D_{\text{geo}}$ converge to similar values for small $\beta$, and the median of $T_{10\%}$ for $D_{\text{pow}}$ is always smaller than for the case of $D_{\text{geo}}$, indicating that the infection takes less time to reach $10\%$ of the vertices in the $D_{\text{pow}}$ case. Conversely, due to the shape of the distribution (Fig.~\ref{fig5}c), the mean value gives a biased value and suggests the opposite effect (Fig.~\ref{fig5}a). To reach $90\%$ of the vertices, however, the situation is different and the two statistics (the median and mean) are similar for both the case of $D_{\text{pow}}$ and for $D_{\text{geo}}$ (Fig.~\ref{fig5}b). At this later stage of infection, the behavior of the epidemics is more homogeneous and we identify a characteristic time in which $90\%$ of the vertices are infected. (Fig.~\ref{fig5}d). These results indicate a slowdown on the infection growth in case of power-law inter-event time distribution (Fig.~\ref{fig5}b), a result that is in agreement with previous theoretical studies~\cite{VRL07,IE09}.

As discussed above, the $D_{\text{pow}}$ case results in a broad distribution of times for early infections ($T_{10\%}$), meaning that although there is a peak in Fig.~\ref{fig5}c, there are several possible scenarios of outbreaks, and these possible scenarios are mainly because of the heterogeneous vertex activity. In fact, out of $1,000$ trials, $145$ result in times above $200$ time steps, and the largest one is $4,453$ time steps. As a comparison, the largest time observed in the $D_{\text{geo}}$ case is $42$ time steps. This narrow distribution of the $D_{\text{geo}}$ cases shows that the choice of the initial condition has little influence on the propagation of the infection. The large $T_{10\%}$ values observed in the $D_{\text{pow}}$ case are responsible for the divergence between the mean and the median. A similar effect, i.e.\ the distribution of inter-event times for the $D_{\text{pow}}$ case being broader than for the $D_{\text{geo}}$ case, is observed when $90\%$ of the vertices are infected. This is not that strong however because for $T_{90\%}$ the distribution is more symmetric in case of $D_{\text{pow}}$ (Fig.~\ref{fig5}d). Out of $1,000$ trials, $71$ result in times above $800$ time steps, being $4,843$ the largest value. As a comparison, the largest value for the $D_{\text{geo}}$ case is $60$ time steps.

\subsection{Effects of the inter-event time on the SII dynamics}

\begin{figure}
\centering
\includegraphics[scale=0.6]{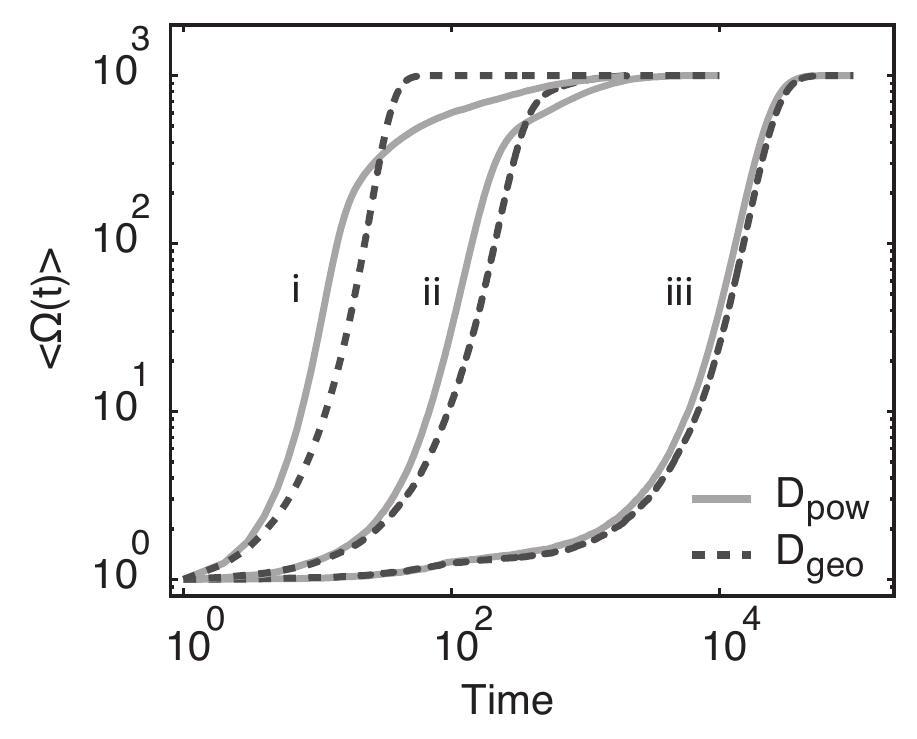}
\caption{Average number of infected vertices $\langle \Omega (t) \rangle$ in case of SII dynamics for different combinations of per-contact infection probabilities $\beta_1$ and $\beta_2$; (i) $\beta_1=1$ and $\beta_2=0.1$, (ii) $\beta_1=0.1$ and $\beta_2=0.01$, and (iii) $\beta_1=0.01$ and $\beta_2=0.001$. The axes are in log-scale.}
\label{fig6}
\end{figure}

In the case of the SII dynamics, vertices go through two infective stages where the individual contribution to the global epidemics varies at each stage. Since we set the period of the first infective stage to $T_1=90$, the growth curve for $\beta_1=1$ and $\beta_2=0.1$ (Fig.~\ref{fig6}) is not much affected in the early stage because most of the network is infected before the period $T_1$. More specifically, according to Fig.~\ref{fig5}b, the mean value to reach $90\%$ of the network is at about $400$ time steps. After the inflexion point, however, we see that the slowdown interval starts earlier in the SII epidemics. For smaller $\beta$ the high and low infective regimes affect the infection growth significantly, especially, for $\beta_1=0.01$ and $\beta_2=0.001$. This effect is particularly interesting because it corresponds to the range of per-contact infection probability of HIV on several societies~\cite{KJW97}. For these values, the epidemics waits roughly $10$ times longer to take off in comparison to a one infective stage scenario  ($T\sim 100$ for SI, and $T\sim 1,000$ for SII, see also Fig.~\ref{fig3}a).

\section{Conclusion}
\label{concl}

Recent research has suggested that temporal constraints on network structures affect the spread of infections. In particular, there is evidence that the time between two consecutive events of vertex activation, observed in some empirical networks, speeds up (or slows down in case of SI-like models of information spreading) the spread of infections in comparison to homogeneous vertex inter-event times. These results are based on contrasting the infection growth on the original empirical network and on null models of networks where the original time-stamps are randomized by a reshuffling procedure. To investigate the changes in the shape of the infection growth more carefully, we propose in this chapter a simple and intuitive stochastic model of a temporal network where the topology is left completely random but the vertex inter-event time is explicitly controlled by using a power-law (with cutoff) and a geometric inter-event time distributions.

By simulating a SI epidemics model co-evolving with the network structure, we observe that the inter-event time affects the infection growth differently according to the stage of the epidemics. On early stages, in comparison to the homogeneous scenario (Poisson-like dynamics), the heterogeneous (power-law) inter-event time results in a faster infection growth, while at later stages, the infection growth is slower. In our model, these differences results from the inter-event time and do not depend on the network topology (e.g.\ degree distribution, community structure, degree-degree correlations). We also observe that the heterogeneous inter-event time creates a diversity of outbreaks and highlights the influence of the initial conditions on the infection spread. In other words, while the homogeneous inter-event time results in characteristic times to reach a certain number of vertices, the heterogeneous produces several scenarios where the infection may take long times before affecting the network significantly.

\section{Acknowledgements}
LECR is beneficiary of a FSR incoming post-doctoral fellowship of the Acad\'emie universitaire Louvain, co-funded by the Marie Curie Actions of the European Commission. AD is a Research Fellow with the Fonds National de la Recherche Scientifique (FRS-FNRS). Computational resources have been provided by the supercomputing facilities of the Universit\'e catholique de Louvain (CISM/UCL) and the Consortium des \'Equipements de Calcul Intensif en F\'ed\'eration Wallonie Bruxelles (CECI) funded by FRS-FNRS. 

%



\printindex
\end{document}